# A digital perspective on the role of a stemma in material-philological transmission studies

KATARZYNA ANNA KAPITAN

Taking its point of departure in the recent developments in the field of digital humanities and the increasing automatisation of scholarly workflows, this study explores the implications of digital approaches to textual traditions for the broader field of textual scholarship. It argues that the relative simplicity of creating computer-generated stemmas allows us to view the stemma codicum as a research tool rather than the final product of our scholarly investigation. Using the Old Norse saga of Hrómundur as a case study, this article demonstrates that stemmas can serve as a starting point for exploring textual traditions further. In doing so, they enable us to address research questions that otherwise remain unanswered. The article is accompanied by datasets used to generate stemmas for the *Hrómundar saga* tradition as well as two custom Python scripts. The scripts are designed to convert XML-based textual data, encoded according to the TEI Guidelines, into the input format used for the analysis in the PHYLIP package to generate unrooted trees of relationships between texts.

## 1. Introduction[1]

The *stemma codicum* — the final product of thorough and painstaking textual examination, the main objective of stemmatology, and the subject of severe criticism from material philologists — has played a central role in textual scholarship for over two centuries. Stemmas enable scholars to illustrate transmissions of texts preserved in multiple manuscripts, reconstruct lost archetypes of literary works, and reveal the best witness(es) of a tradition.[2] The mere fact that an entire branch of philology, revealingly called *stemmatics* or *stemmatology*, is devoted to making stemmas testifies to its prominent role in scholarship.

Despite its importance, when one compares the existing definitions of the term *stemmatics* in the English language, the objective of stemmatics is not straightforward. In the twelfth edition of the *Concise Oxford English Dictionary*, 'stemmatics' is defined as "the branch of study concerned with analysing the relationship between surviving variant versions of a text, especially so as to reconstruct a lost original" (Stevenson and Waite 2011: 1414). Meanwhile, *Oxford English Dictionary* explains 'stemmatology' as "the study which attempts to reconstruct the tradition of the transmission of a text or texts (esp. in manuscript form) on the basis of the relationships between the readings of the various surviving witnesses" (OED, 'stemmatology').[3]

---

[1] This article is partially based on a lecture I delivered in the lecture series of Selskab for Nordisk filologi in 2019. I would like to thank Anne Mette Hansen and Seán D. Vrieland for inviting me to deliver this lecture as well as for the encouragement to submit an article based on it to *Studier i Nordisk*, the editorial board of which I also joined in the same year. I would also like to thank my colleagues at the editorial board and the anonymous peer reviewer for the valuable feedback they provided on the contents of the present article, as well as Böðvar Ólafsson for his feedback on my Python scripts.

[2] These various functions of a stemma have been neatly summarised by Varvaro (1999: 50): "Until then [Giorgio Pasquali's work in the 20s and 30s of the twentieth century] the *stemma codicum* was taken to be an instrument for determining and for making explicit the formula which afforded a mechanical reconstruction of the original; for Pasquali it becomes instead simply a summary of the history of the tradition of a text, a history subject to the editor's investigation."

[3] All references to the *Oxford English Dictionary* are to its online version, www.oed.com, last accessed 3 October 2022. The current online entry is based on the second edition of the dictionary from 1989, which was

Although these definitions are similar in essence, from the perspective of the present study there is a substantial difference worth emphasising. The first definition focuses on the reconstruction of a lost original, while the following one focuses on the transmission of the text and the relationships between the extant manuscripts. So, the process of revealing and understanding the relationships is the objective of the discipline. The dichotomy between these definitions is a result of different methodological approaches to textual traditions. On one side there is traditional textual criticism. On the other, there are theoretical approaches which do not place the archetype as the centre of the examination, such as *Überlieferungsgeschichte* (or simply 'transmission history') and new/material philology. The present study draws extensively on the latter two approaches and explores the role of a stemma in material-philological transmission studies, meaning studies which are not exclusively interested in the original, or the earliest traceable form of a text, but in the changing contexts in which literary texts are preserved. Taking as its point of departure the recent developments in the field of digital humanities and increasing automatisation of scholarly workflows, this study explores the implications that the digital approaches to textual traditions might have for the field of textual scholarship at large. It argues that, thanks to the relative simplicity of creating computer-generated stemmas, we should be able to look at the stemma as a research tool rather than as a final product of our scholarly investigation. We can treat a stemma as a starting point of our explorations of textual traditions and by that focus on research questions that otherwise frequently remain unanswered.

This study consists of two main parts. The first part provides a brief overview of the theoretical background to help the reader understand the context from which the present considerations arose. This overview includes some snapshots from the millennia-long history of textual scholarship, including textual criticism, transmission history, and new and material philology, which lead to a presentation of the happy marriage of the latter two approaches, labelled here as material-philological transmission studies. The second part is devoted to rethinking the role of the stemma in modern textual scholarship, or more accurately in digital textual scholarship. It takes its point of departure in experiments conducted on a seventeenth-century version of a medieval Old Norse work known as *Hrómundar saga Greipssonar* (also *Gripssonar*) using the PHYLIP package (Phylogeny Inference Package; Felsenstein 2013). It briefly outlines the methodology of computer-assisted stemmatics, focusing on one of its possible applications, and then presents a case study illustrating the auto-generated stemmas as an exploratory tool in material-philological transmission studies. It highlights the research questions that could be asked when treating the stemma as an intermediate step in a research process rather than as its final product. The article is accompanied by datasets used to generate stemmas of *Hrómundar saga* as well as two custom scripts that can be used to convert XML-based textual data into the input format used for the analysis in the PHYLIP package.

## 2. Theoretical background

### 2.1 Textual criticism

The essence of textual criticism was described by Paul Maas in his *Textkritik*: "Aufgabe der Textkritik ist Herstellung eines Autograph (Original) möglichst nahkommenden Textes (*constitutio textus*)" (Maas 1960: 5).[4] In this narrow sense *textual criticism* refers to the reconstruction of the text of the archetype, based on the examination of existing witnesses, a practice usually associated with the German philologist and father of eclectic editing Carl Lachmann. The genesis of textual criticism, however, lies in Antiquity. It reaches back to the great libraries of Pergamon and Alexandria, where the principles of scholarly editing were laid down. It is where the two competing schools of textual criticism originate: the school of 'analogy' and the school of 'anomaly'.[5] The school of 'analogy' was developed in Alexandria under the influence of Aristotelian philosophy. Alexandrians searched for an authorial ideal and believed that the original could be reconstructed by collation of variant readings, preserved in corrupted copies, so they used variants to establish a 'correct text'. The school of 'anomaly' was developed in Pergamon under the influence of Stoic philosophy. Pergamonians assumed that the original was not reachable and focused on the language of preserved documents. The legacy of the feud between 'analogists' and 'anomalists' can still be seen in twentieth-century scholarship, as observed by Bloom (1979: 13–14):

> Whereas the analogists of Alexandria held that the literary text was a unity and had a fixed meaning, the anomalists of Pergamon in effect asserted that the literary text was an interplay of differences and had meanings that rose out of those differences. Our latest mimic wars of criticism thus repeat battles fought in the second century B.C. between the followers of Crates of Mallos, Librarian of Pergamon, and the disciples of Aristrachus of Samothrace, Librarian of Alexandria.

Even though Bloom focuses on the poetic theory and literary criticism, not on textual criticism, his observations can relate to textual criticism if we interpret the 'latest mimic wars of criticism' as the clash between best-text editing and eclectic editing. Pergamonians would be seen as the forefathers of best-text editing, while Alexandrians of eclectic editing. Both approaches, however, have the main objective of textual criticism in common: revealing the closest-to-the-original manifestation of a given work.

## *2.2 Transmission history*

The German term *Überlieferungsgeschichte,* here translated simply as 'transmission history', has been also translated as 'history of textual transmission' (Glauser and Tranter 1990: 45–46) and 'textual and transmission history' (Baisch 2017: 178).[6] This approach towards working with pre-modern literature has its origins in the 1960s-70s and has been outlined in the program of the *Würzburger Forschergruppe für deutsche Prosa des Mittelalters*, published in *Jahrbuch für Internationale Germanistik* (Grubmüller *et al.* 1973). In this report, the main objectives of this transmission-historical project were discussed, including the expected results, such as types of modifications present in late-medieval narratives, regional and social variants, and most importantly a contribution to literary history due to the fact that:

---

[4] Maas' publication first appeared in 1927, here it is cited following its 1960 edition. For further discussion of textual criticism see, for example, Thorpe (1972), Reynolds and Wilson (1974), Tanselle (1981, 1992), Greetham (1992, 1994, 2007), Gabler (2005), Timpanaro (2005) and Trovato (2014). In the Nordic context, see Olrik Frederiksen (1991, 2003), Kondrup (2011), Haugen (2013).

[5] For a discussion of the school of Alexandria and Pergamon see Sandys (1903: I, 105–166) and Greetham (1994: 297–302).

[6] Also known as 'transmission-historical method' from German 'überlieferungsgeschichtliche Methode' (Baisch 2017). This appears to be misleading usage, as *Überlieferungsgeschichte* should be seen more as a theory than a method. If we want to look at it as a method, then it must be a method which is a sum of all the methods applied in 'überlieferungsgeschichtliche' research, for example, stemmatology and methods of descriptive bibliography, see the discussion presented by Löser (2016).

> Dadurch, daß der Ansatz nicht beim isolierten „authentischen" Werk, auch nicht beim Autor erfolgt, sondern das Leben des Textes im Spannungsfeld des Autors, der Bearbeiter, der vermittelnden Schreiber und Drucker sowie des rezipierenden Publikums editorisch ausgefaltet wird, ergeben sich Perspektiven einer „historischen" Literaturgeschichte, die sich weitgehend von einer „Denkmäler"-Literaturgeschichte abhebt. (Grubmüller et al. 1973: 171-2)

The metaphor of 'the life of the text' summarises perfectly the main interest and objectives of research within this theoretical framework. The analysis of a literary work within the broad context of its transmission and dissemination, which takes into account the interplay between authors, scribes, and the audience, as well as an interest in variations and modifications made to the text, is an obvious break with traditional textual scholarship. At the same time the group can be seen as a sort of continuation of traditional philology, since it utilises traditional methods, such as stemmatology (Grubmüller et al. 1973: 171). The theoretical framework of *Überlieferungsgeschichte* was used in the volume *Überlieferungsgeschichtliche Prosaforschung*, edited by Kurt Ruh, where an important aspect of the textual tradition of a literary work, a mutation or a reworking, is brought forward:

> Freilich ist dabei nicht nur die Zahl der Überlieferungsträger maßgeblich, sondern die Textmutationen als solche. Es gibt Texte, die sich kaum verändern wie etwa Übertragungen biblischer Bücher, aber auch, generell, poetische Texte; andere, die vielfältigem Gebrauch unterliegen, erscheinen in den verschiedensten Gestalten und Ausformungen, werden „durchschossen" von anderen Texten oder in andere inseriert, erweitert, gekürzt, purifiziert (etwa hinsichtlich obszöner oder häretischer Partien), versifiziert oder in Prosa umgeschrieben: alles nur Erdenkbare ist textgeschichtliche Wirklichkeit. (Ruh 1985b: 268–269)

Ruh observed that everything that happens to the text, including prosification or versification of a given work, lies within the area of interest for transmission historians. Therefore, we can propose that a study of the transmission history of a given work should take into account not only the transmission of a single work, but also various reworkings or adaptations of that work. By studying reworkings or adaptations of a given work in other media, transmission historians are creating an interdisciplinary field of interplay between textual criticism, literary studies, and history. They are still interested in revealing the archetype of a given tradition (textual criticism), but this is only a starting point of the analysis, because they are more interested in the changes that arose in the course of the transmission and the adaptations of a given work (literary studies) and what these changes and adaptations can tell us about the society that produced these texts (history).

### 2.3 New and material philology

New philology and material philology are theoretical frameworks which have become very popular at the tun of the twentieth and twenty-first centuries. They appear in various guises and under many names, for example 'artefactual philology', emphasising the importance of a codex 'both as the material embodiment of the text and as cultural artefact in its own right' (Hansen 2004b: 239),[7] 'renewed philology' (Vitz 1993; Williams 2009), and 'descriptive philology' (Bäckvall 2013: 48–50; 2017: 29–30). Even though the manifesto of new philology is traditionally associated with the 1990 volume of *Speculum* and Nichols's introduction to the volume, already in the *Romanic Review* from 1988 (Nichols 1988) three essays are classified under the common header 'Text and Manuscript: The New Philology'.[8] This theoretical approach of current new philology has its roots in the Romance scholarship of the 1970s and 1980s, as clearly outlined by Hansen (2006: 93–94), including, for example, Cerquiglini's (1989) *Eloge de la variante, histoire critique de la philologie* (English translation from 1999) and Zumthor's (1972) idea of 'mouvance' as discussed in *Essai de poétique médiévale* (English translation from 1992).[9]

---

[7] Artefactual or material philological approach has played an important role in studies by Anne Mette Hansen, for example Hansen (2004a, 2012, 2017). The term 'artefactual philology' in English first appeared in Driscoll (2010).

The main point of new philology, as introduced in 1990, was a revival of philology with its new, variant- and material-oriented flavour, which stands in opposition to the 'old' philology of fixed texts, as Nichols writes in his introduction to the volume:

> [P]hilology represented a technological scholarship made possible by a print culture. It joined forces with the mechanical press in a movement away from the multiplicity and variance of a manuscript culture, thereby rejecting, at the same time, the representation of the past which went along with medieval manuscript culture: adaptation or *translatio*, the continual rewriting of past works in a variety of versions [...]. The high calling of philology sought a fixed text as transparent as possible, one that would provide the vehicle for scholarly endeavour but, once the work of editing accomplished, not the focus of inquiry. It required, in short, a printed text. (Nichols 1990: 2–3)

New philology recognises the instability of the medieval text; it turns towards the origins of medieval literature and meets it where it is, in the multidimensional matrix of manuscripts:

> The apparently straightforward act of copying manuscripts is not free from mimetic intervention, either. In the act of copying a text, the scribe supplants the original poet, often changing words or narrative order, suppressing or shortening some sections, while interpolating new material in others. As with the visual interpolations, the scribal reworkings may be the result of changing aesthetic tastes in the period between the original text production and the copying. (Nichols 1990: 8)

The material aspects of the new philological approach were emphasised by Nichols in 1994 when the term new philology was substituted by material philology. While in the 1990 essay, textual variance and materiality can be seen as two equally important focal points of new philology, in 1994 we can see a clear shift in focus from the text to the artefact, when Nichols wrote: "The philology of the manuscript or material philology must focus on the dynamics of the expressive systems within the manuscript viewed as cultural artefact" (Nichols, 1994: 118). Nichols further developed his ideas in 1997 when he argued that a manuscript should be seen as the manifestation of the historical context in which medieval literature was produced and in which it should be interpreted (Nichols 1997: 10–11). He focused on the multi-dimensional nature of the manuscript matrix, and emphasised that:

> [T]he manuscript space evokes a variety of historical issues, such as the relationship between picture and text, context and rubric [...], but also broader questions having to do with the representation of a given work or set of works in a particular place and moment. [...] [T]he manuscript is not simply, or not at all a vehicle for conveying a literary work; rather it represents a culture in which the literary work is one among many components, primus inter pares, perhaps, but not preeminent. (Nichols 1997: 14)

---

[8] Some critics of new philology, usually those who also like to emphasise that there is nothing new about new philology, emphasise that this and similar terms were appearing throughout centuries, for example, in Barbi's *Nuova filologia* from 1938, or Vico's *Scienza Nuova* from 1725, cf. (Bloch 1990; Busby 1993; Wolf 1993; Varvaro, 1999; Sverrir Tómasson 2002).

[9] To the best of my knowledge, Anne Mette Hansen (2006), was the first scholar to observe Zumthor's influence on new-philological thought and this line of interpretation was followed by Driscoll (2010). The immediate response to the new-philological postulates of Cerquiglini and Nichols can be found in *Towards a Synthesis? Essays on the New Philology?* (Busby 1993) which gives insight into the heated atmosphere of the late 1980s and early 1990s. The criticism of the new-philological approach in Old Norse studies can be found, for example, in works by (Wolf 1993; Sverrir Tómasson 2002; Haugen 2010). More recently Westra (2014: 15) presented New Philology as an ideological construct and Material Philology as an actual practice of working with texts and manuscripts. Other recent expressions of criticism have been brought forward by Males (2023) and Myrvoll (2013), the latter with an anecdotally brief reference list.

The focus on materiality allowed more of an artefactual approach to manuscripts, not only as text-bearers, but also as physical objects which were expressions of cultural production.

### 2.4 Material-philological transmission studies

Both *Überlieferungsgeschichte* and new/material philology can be seen as a response to the traditional text-critical approach and its prioritization of, if not the original, then what is closest to the original. Both *Überlieferungsgeschichte* and new/material philology emphasise the variation and the instability of a medieval text and both approaches are more or less manuscript-oriented. They are, however, distinct and in many aspects can be seen as complementary to one another.[10]

Material philology, in its purest form, takes the artefact as its point of departure rather than the text. It draws on the methods of, for example, bibliography and the history of the book to interpret a given manuscript in its entirety, including but not limited to the text.[11] Material philology allows us to look at a text in its physical context but it does not give us any tools to establish the relationships between the various texts of the same work preserved in the manuscripts, as it rejects traditional approaches to texts and makes the manuscript the main point of the analysis. It considers all the textual variants equally interesting, and therefore there is no need to establish any relationships between the extant texts as they are simply not the core subject of the study.[12] We take a manuscript as it is and edit its texts, we discuss its physical aspects, such as script, layout and codicology. We might discuss the socio-economic status of the scribe, if we are interested in the sociology of texts[13] and social dimensions of literary production, but we do not engage in the stemmatic discussion of the exemplar, copy, sibling etc.

The 'überlieferungsgeschichtliche' approach allows us to look at the changes in the text in time and space. We know how the subsequent texts of a given work relate to each other, because we employ the methods of textual criticism, such as the stemmatic method, but we do not necessarily look in depth at the manuscripts' physicality. The main focus is literary: it lies in the textual manifestations of a given work, and the manuscript as an artefact plays a secondary role. The importance of variation and the instability of the text resembles somewhat the postulates of new philology as introduced in 1990, but it is clear that there is a big difference between material philology and *Überlieferungsgeschichte*. That difference lies in the manuscript itself, as Ruh (1985a: 267) writes: "Es geht hier indes nicht um die Faszination der Handschrift, so willkommen sie als Anreiz ist, sondern um ihre spezifischen Funktionen in der Konstitution der Geschichtlichkeit von Literatur". This focus on the history of literature rather than on artefacts resulted in a split from what in 1990 might have looked like two sides of the same coin. The practitioners of *Überlieferungsgeschichte* remained textual and carried on with their synoptic editions of various versions of the same work, while the material-oriented researchers went a step further, envisioning, for example, editions of full manuscripts, not individual texts preserved in a given manuscript.[14]

---

[10] As Anne Mette Hansen (2012: 9) puts it "An artefactual philological examination of a manuscripts comprises the whole book, i.e. codicology (including collation, signatures, layout and decoration), textual analysis, the relationship between text and images, and script/orthography." Quite naturally then, material philological studies take as their point of departure single codices, rather than literary works, and analyse them according to the described criteria.

[11] For an introduction to history of the books, see, for example Gaskell (1972), Eliot and Rose (2007), there further literature.

[12] Some scholars emphasise that new philology is secondary to *Überlieferungsgeschichte*, for example Williams-Krapp (2000) and more recently Löser (2016: 6), who emphasises that *Überlieferungsgeschichte* puts the variant at the centre of research long before Cerquiglini. It must be noted that I have failed to observe any references to *Überlieferungsgeschichte* in the 1990 *Speculum* volume, and the field of expertise of scholars who published in this volume was French and English literature, not German literature, so there was most likely no dialogue between these two trends. This does not mean, however, that there is nothing new in new philology, since the *Überlieferungsgeschichte* cannot answer, or is not interested in, all the questions that are important from the material-philological perspective (cf. Baisch 2017).

[13] For an introduction to sociology of texts, see, for example McGann (1983) and McKenzie (1986).

Neither approach alone is sufficient to gain a holistic overview of transmission histories of texts existing in multiple versions and preserved in multiple manuscripts, but a happy marriage of both can remarkably benefit our understandings of texts in contexts. The contexts in which texts circulate include:

- the intertextual context, understood here as the relationship between various adaptations of a given story,
- and the material context, understood as the physical environment in which a given story appears, including other texts appearing in the same manuscripts.

The application of material philological approaches to the artefacts, supported by systematic analysis of the textual transmission, established by the application of genealogical criteria, is, in my view, most beneficial for expanding our knowledge of transmission and reception of literary texts in the past.[15] Such material-philological transmission studies put equal emphasis on the transmission of texts and the materiality of artefacts. While there is no one-size-fits-all methodology in philology, as every research project has its own questions and requires different approaches to the source material, the material-philological and transmission-historical approach enables us to provide the most holistic view of the history of texts.

## 3. Stemmatics in the digital age

### 3.1 Computer-assisted stemmatics in material-philological transmission studies
Material-philological transmission studies do not aim to reconstruct the archetype or reveal the best text of the tradition, but still, they apply the principles of stemmatics to reveal the relationships among the extant texts. Therefore, the function of the stemma is different than in traditional textual criticism. It is not a tool for establishing the text by the application of the majority rule. It is used to visualise the relationships between the texts and deliver a point of departure for further investigation. Unavoidably, as a side product, or an additional benefit of the applied methodology, the best text for a given tradition will most likely be revealed, but it is not the objective of the study in its own right. It is a point of departure for the analysis of the transmission of the text(s), of the changes introduced to them, and of the material contexts in which they are preserved. These are the main objectives of a study, and a stemma is only one of the tools used to achieve them.

---

[14] It should be noted here that in Old Norse studies it has been a practice to edit important manuscripts in their entirety long before material-philological turn, for example Flateyjarbók (Guðbrandur Vigfússon and Unger 1860–1868), Codex Frisianus (Unger 1871), Hauksbók (Eiríkur Jónsson 1892), Landnámabók (Finnur Jónsson 1900). These editions, however, do not situate material-philological values as the main focus of the investigation; they are mainly textually and linguistically oriented. Also, more recent publications, such as editions of Möðruvallabók, AM 132 fol. (de Leeuw van Weenen 1987) and Alexanders Saga, AM 519a 4to in the Arnamagnæan Collection, Copenhagen (de Leeuw van Weenen 2009), as well as studies of Morkinskinna (Kjeldsen 2013), put far more emphasis on the language of the text than on the materiality of the artefact.

[15] Similar postulates have accompanied new philology and its relationship to 'old' philology" from the early days, as expressed for example by Maddox (1993, 61), who asked the important question of whether a philologist must choose between 'old' and 'new' instead of getting best of both words, and more recently by Leslie (2012, 151), who wrote that both approaches, 'old' and 'new philology' can be complementary if one is prepared to negotiate.

This postulate might raise some objections, as the mere preparation of a stemma is a large enough task for any research project. Therefore, it might not seem realistic to make it an intermediary step in a research project that has totally different objectives. This is only partially true in the digital age, where computer assisted methods allow us to automate many tasks, including revealing and visualising relationships between texts. With the rise of computer-assisted stemmatics, new possibilities for engaging with textual traditions have appeared. Due to a lack of space, the present study cannot provide an overview of the history of computer assisted stemmatics, but fortunately an up-to-date overview of the discipline can be found in the recently published *Handbook of stemmatology* (Roelli 2020b). It can serve as a useful introduction to the method, its history and applications.

The handbook also provides plenty of valuable insights into the role and value of the stemma, emphasising its importance in textual scholarship. Roelli writes in his chapter entitled "Definition of stemma and archetype" that *stemma codicum* in Latin simply means "genealogical tree of the manuscripts" (Roelli 2020a: 210) and that in practice a stemma is "an oriented tree-like graph representing a hypothesis about genealogical relationships between witnesses of a text" (Roelli 2020a: 212). Even though Roelli himself notes that philological stemmas are not always trees *sensu stricto*, for example due to contamination, there is a distinction between a stemma, which is a result of philological investigation, and a graph (or an unrooted tree), which is a result of computer-assisted analysis. This distinction is maintained in other contributions to the *Handbook*, including these by Guillaumin (2020), Macé (2020), and Roos (2020). As van Zundert (2020: 293) puts it in his introductory remarks, the graphs (or unrooted trees) can be merely regarded as "hypotheses for stemmata". Creating a hypothesis for a hypothesis seems quite superfluous. If a stemma itself is a hypothesis, is there then a need for yet another hypothesis? It is, however, understandable why scholars tend apply this terminological differentiation between a graph and a stemma. A graph is a result of computer-assisted analysis based exclusively on textual readings and it can be unrooted and undirected. A stemma is a rooted and directed graph which is also based on textual readings, but it, furthermore, takes into account other pieces of information, some of which are internal, others external to the manuscripts preserving the text in question. This distinction is, however, not applied in this study, as I refer to a stemma as an exploratory tool and what I have in mind is a collection of graphs or unrooted trees generated for the tradition.

There are many programs that can be used to conduct computer-assisted stemmatic analysis, many of which have been described in the *Handbook* (Hoenen 2020a; Macé 2020). They apply different methodological principles and have various requirements for technical skills. The present study uses the PHYLIP package (Phylogeny Inference Package), which is relatively easy to use and has been for the first time successfully applied to Old Norse tradition by Alaric Hall and Katelin Parsons (2013) in their study of *Konrads saga keisarasonar*.[16] The PHYLIP package was first released in 1980 and it is still maintained (Felsenstein 2013). Among many specialised programmes, it contains three programs that lend themselves especially useful for textual studies: PARS, CONSENSE and DRAWTREE.

For the sake of accessibility of this study, it is useful to briefly describe the workflow of using PHYLIP for those readers unfamiliar with its characteristics. In order to conduct computer-assisted stemmatic analysis with the PHYLIP package, variants appearing in the tradition have to be converted into numeric format. PARS, one of the programmes included in PHYLIP, requires numeric values, which represent shared readings. This corresponds to the DNA sequences that PARS was originally designed to analyse. The process of conversion of readings to numbers can be done in different ways. One way of doing this is to transcribe and collate all the texts in a spreadsheet, then manually assign numeric values to different readings, and finally copy these numbers into a plain text file following the formatting required by PARS. This is a method that was used in Hall and Parsons's (2013) study and in my previous work (Kapitan 2017). From my experience, this method is very time consuming, and it is easy to generate errors. A human being manually assigning numbers to readings in a spreadsheet is not very effective, and, at times, not very consistent. Moreover, transcribing all the texts into a spreadsheet is quite repetitive, as the very same reading may appear in all manuscripts and then the same number has to be manually assigned to all of them.

---

[16] Other studies have since followed, among them Zeevaert *et al.* (2013) and Kapitan (2017).

To simplify this process and limit repetitive tasks to minimum, I have conducted some experiments aimed at developing an alternative workflow.[17] In this approach, instead of transcribing all texts into a spreadsheet, one can use a base-text transcription of a single witness with variant readings encoded in XML, following the guidelines of the Text Encoding Initiative (TEI Consortium 2020). So, the process of preparing an auto-generated stemma starts with a TEI-XML-based transcription of the base text and encoding of variant readings from other witnesses, while it ends with a graphic visualisation of the relationships among the witnesses. The workflow consists of four main steps (visualised in Figure 1):

1. On the first step, variant readings are encoded in the apparatus entry element <app>, which includes a lemma element <lem> and at least one reading element <rdg>. The lemma element contains the main reading, or the reading of the base text. The reading elements contain the variant readings. Both <lem> and <rdg> elements have to contain a witness attribute @wit, which specify which of the manuscripts preserve the selected reading. The portion of text where no textual variation appears is transcribed only once and is not encoded within the apparatus element. The readings shared by multiple manuscripts are transcribed only once and it is indicated through the values of the witness attribute which manuscripts preserve the reading. This saves a significant amount of time.

2. The second step is a transformation with Python, which replaces the process of assigning numeric values manually to the variant readings in the spreadsheet. A custom Python-script extracts information encoded in the XML-file and transforms it into a txt-file, which will be used as an input file for PARS. The Python script finds each <app> element, checks how many children it has, and then assigns the same numeric value to each witness that contains the same reading. As an output it generates a plain text file that can be used as an input for PARS.[18]

3. The third step is PARS processing. The txt output file from our Python-transformation is used as an input file for PARS. PARS generates another output file, which includes the information about the most parsimonious trees of relationships. This output file can be used as an input for the visualization software.

4. So the last, fourth, step is the visualisation and interpretation. The output from PARS is used as input for DRAWTREE to obtain the visualization of the unrooted tree of relationships.

[18] Due to the limitation of PARS, which can process up to eight different characters, my tests focused exclusively on major variants, which allowed me to limit the extent of the variation within the tradition.

In some cases, an additional step can be introduced. If multiple alternative unrooted trees are generated by PARS, we might want to visualise each of them separately, or use the CONSENSE programme, also included in the PHYLIP package, to achieve a consensus tree for the tradition. The visualisations can then be used for manual interpretation by a researcher and, as a result, auto-generated stemmas serve as an exploratory tool for textual traditions. The next section illustrates how this can be applied to real-life examples and how benefit any material-philological transmission study.

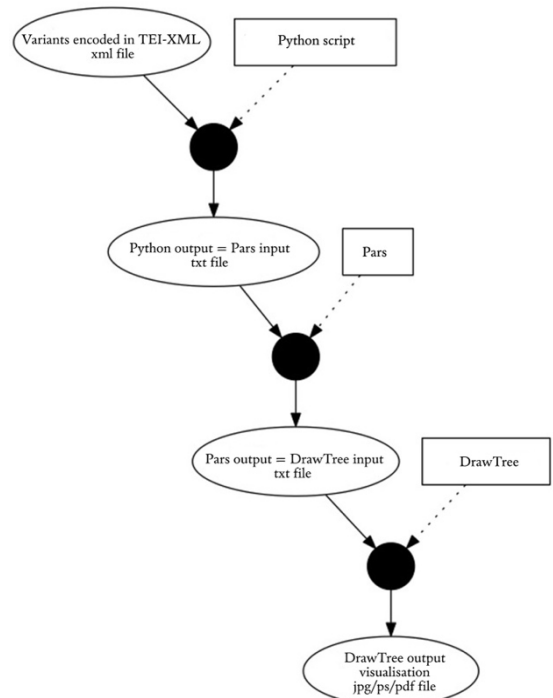

*Figure 1. Workflow: from variants encoded in TEI-XML to an unrooted-tree visualisation.*

### 3.2 Auto-generated stemmas as an exploratory tool

This section demonstrates how auto-generated stemmas (or unrooted trees of relationships, to be precise) can be used as exploratory tools for textual traditions. For this purpose, I will use the case study of the seventeenth-century version of *Hrómundar saga Greipssonar*, which survives in thirty manuscripts.[19] The text has been edited twice: by Erik Julius Björner (1737) and by Carl Christian Rafn (1829–1830), but quite naturally neither of these editions can be considered text-critical by modern standards.[20] Various scholars, however, including Kölbing (1876), Andrews (1911) and Hooper (1930, 1934), have addressed the relationships between some of the extant witnesses of the saga without undertaking editorial work. What is interesting, however, is that they arrived at contradicting conclusions. Moreover, it is worth emphasising that the previous research has been exclusively focused on the oldest part of the tradition, including some six manuscripts, and ignoring the remaining witnesses. There is no evidence in printed scholarship for any of the younger manuscripts ever being examined.

    Following the fundamental rule of textual criticism, *recentiores, non deteriores*, which seem to flourish under the influence of new/material philology and transmission history in studies such as Lavender's (2014, 2020) examination of *Illuga saga*, I also decided to thoroughly examine all known texts of *Hrómundar saga* preserved in extant manuscripts.

---

[19] On the transmission history of *Hrómundar saga*, see Kapitan 2018, 2021a, 2021b, 2022, 2024a, 2024b.

[20] Various popular editions of the text survive, but they are all indebted to the text established by C.C. Rafn, among them Valdimar Ásmundarson (1886), Guðni Jónsson and Bjarni Vilhjálmsson (1943–1944), Guðni Jónsson (1954–1959).

Thanks to this, I was able to determine that the saga that appears in catalogues as *Hrómundar saga* encompasses at least two distinct versions, one from the seventeenth century, and another probably from the nineteenth century. The existence of the nineteenth-century version was first brought to the attention of the scholarly community in my recent studies (Kapitan 2021a, 2021b, 2024a, 2024b).[21] The older saga is preserved in thirty manuscripts which are listed below by repository together with the sigla (in parentheses) under which they are referenced in the present study:

- British Library, London
  – BL Add. 4859 (B4859)
  – BL Add. 4875 (B4875)
  – BL Add. 11108 (B11108)
- Den Arnamagnæanske Samling, København
  – Accessoria 61 (Acc61)
- Det Kgl. Bibliotek, København
  – Kall 614 4to (K614)
  – Thott 1768 4to (T1768)
- Héraðsskjalasafn Akureyrarbæjar og Eyjafjarðarsýslu, Akureyri
  – G-52/1 (G52)
- Landsbókasafn Íslands, Reykjavík
  – Lbs 222 fol. (L222)
  – Lbs 381 fol. (L381)
  – Lbs 633 fol. (L633)
  – Lbs 840 4to (L840)
  – Lbs 1217 4to (L1217)
  – Lbs 1767 4to (L1767)
  – Lbs 2316 4to (L2316)
  – Lbs 2943 4to (L2943)
  – Lbs 3164 4to (L3164)
  – Lbs 4825 4to (L4825)
  – Lbs 3795 8vo (L3795)
  – Lbs 4460 8vo (L4460)
  – ÍB 43 fol. (I43)
  – JS 634 4to (J634)
  – JS 102 8vo (J102)
- Kungliga biblioteket, Stockholm
  – Papp. Fol. nr 67 (P67)
- Staatsbibliothek zu Berlin, Preußischer Kulturbesitz, Berlin
  – Ms Germ qu. 27 (M27)
  – Ms Germ qu. 936 (M936)
- Stofnun Árna Magnússonar í íslenskum fræðum, Reykjavík
  – AM 193 e fol. (A193)
  – AM 395 fol. (A395)
  – AM 345 4to (A345)
  – AM 587 b 4to (A587)
  – AM 601 b 4to (A601)

[21] Printed and digital editions of three versions of *Hrómundar saga* were published in 2024, see Kapitan (2024a, 2024b), the seventeenth-century version is available at https://editions.mml.ox.ac.uk/editions/hromundar_A601/, the eighteenth-century version at https://editions.mml.ox.ac.uk/editions/hromundar_J634/, and the nineteenth-century version at https://editions.mml.ox.ac.uk/editions/hromundar_B11109/.

By surveying the entire tradition, I was able to ask questions about the transmission history of the text, including considerations of who copied and transformed the text of the saga, when, and why, as well as in what material contexts the texts of *Hrómundar saga* appear and what other texts are frequently copied together with it. In order to gain a quick overview of the entire tradition, I used computer-assisted methods of data collations and interpretation, described in the previous section. Figure 2 presents one of unrooted trees of relationships among all thirty manuscript witnesses of this tradition. In the rightmost part of the tree, there is a group of manuscripts which appear to be closely related to each other. These are five manuscripts held in Landsbókasafn in Reykjavík, all written in Iceland in the nineteenth century: J102 (JS 102 8vo), L2943 (Lbs 2943 4to), L4460 (Lbs 4460 8vo), L3795 (Lbs 3795 8vo), and L2316 (Lbs 2316 4to). They all appear to be derived from a common ancestor, which is closely related to two texts preserved in seventeenth-century manuscripts, today held in Reykjavík, Stofnun Árna Magnússonar: A587 (AM 587 b 4to) and A193 (AM 193 e fol.).[22] This is interesting, because A587 and A193 were most likely written in Copenhagen and did not make it to Iceland until their 'repatriation' in the late twentieth century. They became a part of the Arnamagnæan Collection in Copenhagen already in the early eighteenth century, when Árni Magnússon obtained them from the widow of Thormodus Torfæus (Kapitan 2022). While it cannot be excluded beyond doubt that some another closely related copy derived from A587 or A193 made its way to Iceland and there became an exemplar for these manuscripts, there is also another possible explanation.

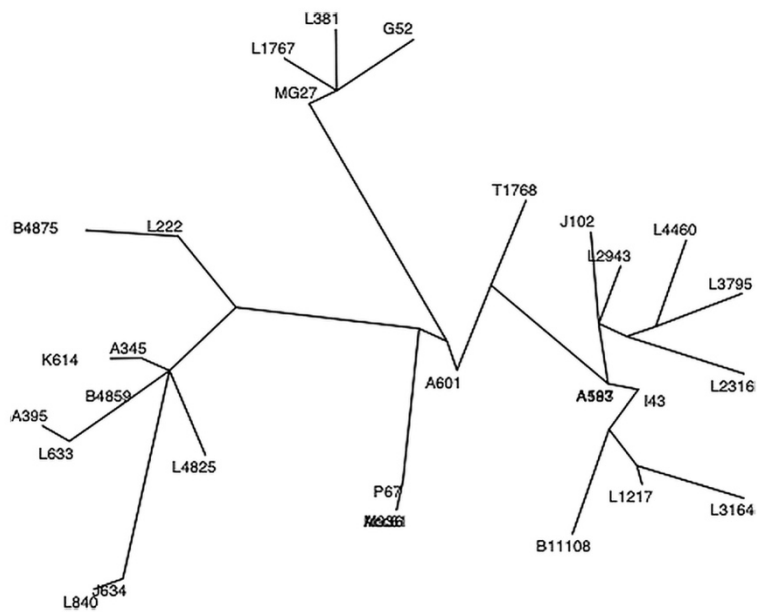

*Figure 2. Unrooted tree of relationships among manuscripts of Hrómundar saga.*

It must be emphasised here that the tree in Figure 2 was built exclusively on the transcriptions of texts preserved in manuscript form, so none of the printed editions was included in the dataset. Rafn's edition from 1829, however, was based on the text of A587. Given the chronology of these manuscripts, all post-dating the publication of the printed edition, and knowing the limitations of programmes such as PHYLIP, discussed recently by, among others, Hoenen (2020b) Guillaumin (2020) and Roos (2020), we can place a hypothesis that all the five manuscripts are derived from Rafn's edition.

A close textual analysis of this group demonstrated that these manuscripts are the only ones that reproduce the major error of Rafn's edition. In A587 and A193 one of the characters is called Kára, but the editor misread the word-initial 'K' as 'L'. This led to the corrupted name 'Lára' instead of 'Kára'. Moreover, all these manuscripts divide the saga into ten chapters, a feature which was introduced to the *Hrómundar saga* tradition by Björner and reproduced by Rafn. However, A587 and A193 divide the saga into five chapters. Moreover, when the readings from the printed editions are included in the dataset, Rafn's edition appears in close proximity to this manuscript group. This is visualised in Figure 3, which presents another unrooted tree of relationships, generated with a dataset that included variants from the printed editions.

Moving to another area of the unrooted tree in Figure 2, in its lower part, there are two manuscripts which appear to be derived directly from a seventeenth-century manuscript held today in the Kungliga biblioteket in Stockholm: P67 (Papp. fol. nr 67). These are two nineteenth century manuscripts: M936 (Ms Germ qu. 936), written in Germany, and Acc61 (Accessoria 61), written in Denmark.[23] P67, from which these two manuscripts are derived, was in a Swedish collection from the late seventeenth-century. Therefore, it is quite intriguing how two manuscripts written in the nineteenth century, one originating in Germany and the other in Denmark, preserve texts of *Hrómundar saga* which are so similar that from the text-critical point of view they are almost identical. The answer lies again in a printed edition, this time in Björner's edition, published in 1737. Björner published his edition in Sweden, and not surprisingly based his text on the only manuscript of *Hrómundar saga* that was available to him in a Swedish repository, P67. He reproduced the text of P67 fairly reliably, with only few minor errors, which made their way into both Acc61 and M936, revealing their clear dependence on the edition. Finally, just as with the previous example, when the readings from the printed editions are included in the input file for PARS, the relationship between the texts preserved in manuscripts and printed editions is recognised by the software (Figure 3).

---

[23] The overlapping layers in Figure 2 indicate that the data encoded for these two texts are identical, which makes it difficult to see that the signatures are M936 and Acc61.

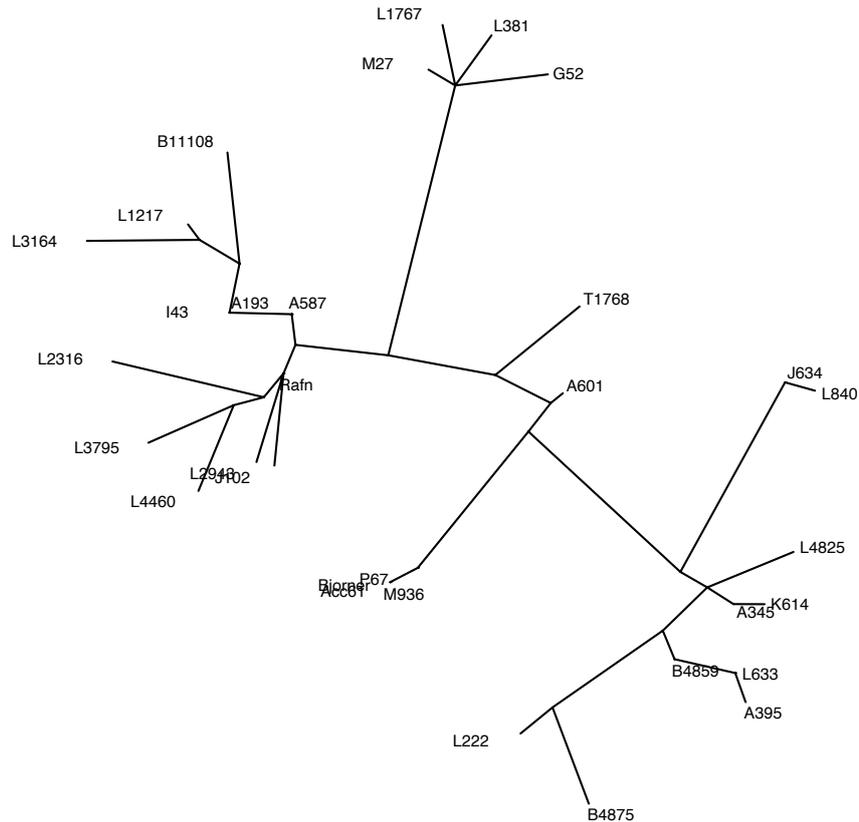

*Figure 3. Unrooted tree of relationships among manuscripts of Hrómundar saga and its two earliest printed editions.*

These two groups create a fascinating case study of manuscripts that preserve texts derived from printed editions. There are five Icelandic manuscripts from the nineteenth century based on Rafn's edition, and another two nineteenth-century manuscripts, one Danish and one German, based on Björner's edition. This allows us to ask questions about how the scribes treated the text of the edition. Did the printed text have any authority over a hand-written text, meaning did the scribes copy the printed text more reliably than when they copied from another manuscript? What kind of changes, if any, are introduced to the text? Some of these questions have been discussed elsewhere, but overall, we can say that the texts written by non-Icelanders were much more faithful to their sources than the texts copied by Icelanders.[24] The variation in non-Icelandic manuscripts is only on the word level, involving misreading of certain characters, while the variation in Icelandic manuscripts involves rewriting, including adding and deleting whole sentences. Also, the manuscript context in which *Hrómundar saga Greipssonar* appears is very different. While the Icelandic manuscripts appear to be free compilations of various sagas, from the sagas of Icelanders, through legendary sagas to romances, the non-Icelandic manuscripts preserve exclusively legendary sagas, and moreover they preserve them in more or less the same order in which they appear in the printed edition.

Finally, there is one more group of manuscripts that will serve as my last example for the usage of a stemma as exploratory tool. It consists of three manuscripts: B4859 (BL Add. 4859), written ca. 1695, L633 (Lbs 633 fol.); written before 1721; A395 (AM 395 fol.), written ca. 1764 (on the left of <mark>Figure 2</mark> and right of <mark>Figure 3</mark>). The textual connection between these manuscripts can be explained by the analysis of the scribal networks in Northwest Iceland, partially influenced by family relations. <mark>Figure 4</mark> illustrates the locations associated with these manuscripts, blue markers for B4859, green for L633, and yellow for A395. Vigur is where the commissioner of B4859, Magnús Jónsson, was living. After his death in 1702 at least some of his manuscripts were inherited by his daughter. B4859 had to be one of them, as it appears on the list of books owned by her husband. They lived in Viðidalstunga.

---

[24] For an in-depth study of the manuscripts derived from printed editions, see Kapitan (2019, 2021c).

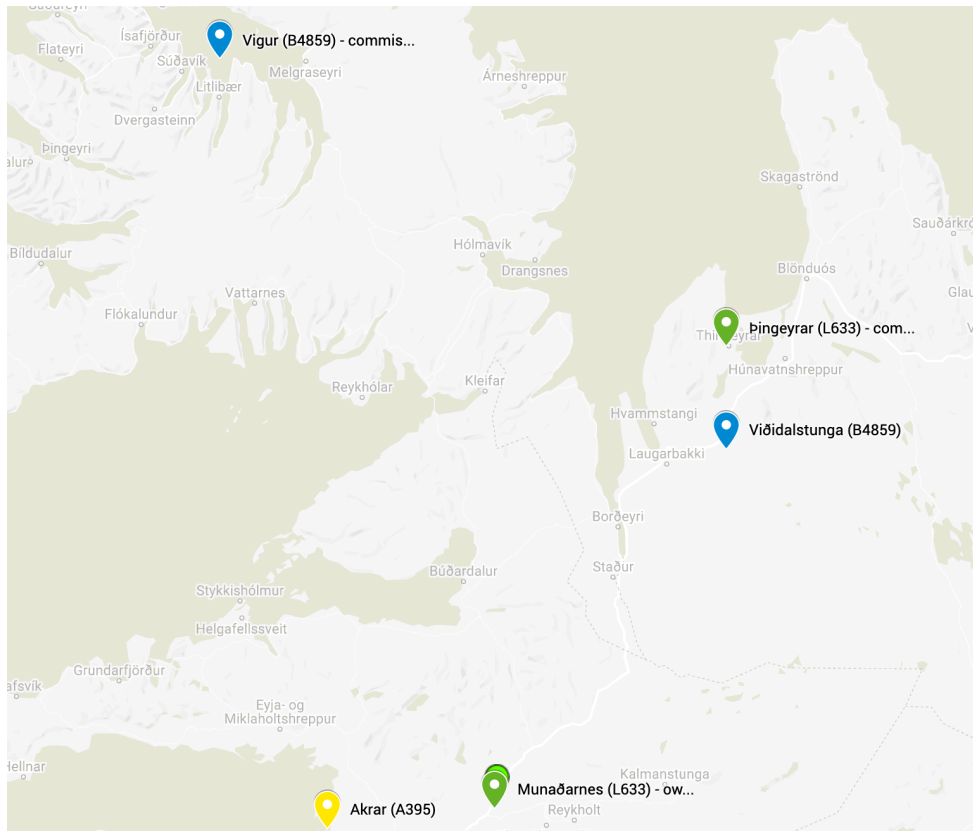

*Figure 4. Locations associated with three manuscripts of Hrómundar saga.*

While B4859 was in Viðidalstunga, Lauritz Gottrup commissioned the writing of L633 in Þingeyrar, ca. 25 km to the north. The whereabouts of L633 after Gottrup's death in 1721 are unknown, but around 1750 the manuscript was owned by Kár Ólafsson from Munaðarnes. Finally, in the 1760s, A395 was written in Akrar, around 50 km west of Munaðarnes. Given this geographic proximity of an assumed exemplar and its copies, it seems highly possible that in these manuscripts the texts of *Hrómundar saga* were copied one after another.

Interestingly, in addition to *Hrómundar saga*, there are some other texts that appear in all three manuscripts. These are *Bragða-Ölvis saga, Mírmanns saga*, *Hervarar saga og Heiðreks,* and *Kirjalax saga.* While most of these sagas still await their full transmission historical studies, the study of the transmission history of *Mírmanns saga* by Desmond Slay has demonstrated that *Mírmanns saga's* texts in these three manuscripts can be organised in an ancestor-descendant order (Slay 1997). This is also the case with *Hrómundar saga.* It would be beneficial to know whether the remaining texts preserved in these manuscripts are also textually closely related. This is especially true for the case for *Braga-Ölvis saga*, which appears together with *Hrómundar saga* in Reykjavík, Stofnun Árna Magnússonar AM 601 b 4to, which preserves the best and oldest texts of these two sagas (Hooper 1932a, 1932b; Teresa Dröfn F. Njarðvík 2019). Is it possible that all these texts were copied one from another when these manuscripts circulated in the North-West Iceland at the beginning of the eighteenth century? It seems to be the case, but more research is needed to provide convincing evidence.

## 4. Conclusions

In this study, I demonstrated how auto-generated stemmas (or more precisely unrooted trees resulting from computer-assisted analysis) can serve as exploratory tools for textual transmissions and lead to new discoveries, beyond the mere revealing of textual relationships. Having access to multiple versions of the unrooted trees, generated with different criteria and algorithms, allows researchers to identify groups of texts that are closely related and then focus on them to peruse other research questions. These research questions can concern, for example, the nature of variation appearing in the tradition, the significance of local scribal networks in works' transmission, or the material and intertextual contexts in which works appear – so, in practice, the types of questions that go way beyond the objectives of the traditional textual criticism. These are questions that could be asked by not only by practitioners of transmission history, new or material philology, and sociology of texts, but also historians of ideas, or historians of literature. Thanks to the application of computer-assisted tools, the stemma (or its unrooted counterpart) is a useful tool that can benefit various types of interdisciplinary studies.

There are three clear advantages of the application the method described in the present study that I want to emphasise here. First, the automatization of the process of variant collation and calculation of distances among texts makes the process more time-effective than manual data collection and interpretation. What used to take years, to manually collate and evaluate variants, now can be done significantly faster. This allows us to use time for actual interpretation of variants in broader context and ask further questions about textual transmission. Second, the use of the computer to convert the readings into numbers and to calculate the distances between texts minimises the risk of human error and inconsistencies in our data set. Finally, by encoding our variants in XML, we can not only generate a digital edition with an extensive variant apparatus, but also utilise the already encoded information as a dataset for computer-assisted analysis. The information about the relationships among manuscripts is already encoded in our edition, so various unrooted trees can be generated using different criteria. For example, we can prepare separate trees for each section of a work or base our visualisation on different types of variants. This makes our research not only reproducible, but also gives the user of our digital edition possibility to explore the textual tradition in his or her own way, making the edition itself a research tool.

## Bibliography

Andrews, Albert LeRoy (1911) Studies in the Fornaldarsǫgur Norðrlanda. *Modern Philology* 8: 527–544. https://doi.org/10.1086/386846

Bäckvall, Maja (2013) *Skriva fel och läsa rätt? Eddiska dikter i Uppsalaeddan ur ett avsändar- och mottagarperspektiv*. Uppsala: Uppsala universitet, Institutionen för nordiska språk.

Bäckvall, Maja (2017) Description and Reconstruction: An Alternative Categorization of Philological Approaches. In Harry Lönnroth (ed.) *Philology Matters!: Essays on the Art of Reading Slowly*: 21–34. Leiden and Boston: Brill. https://doi.org/10.1163/9789004349568_003

Baisch, Martin (2017) Transmission and Materiality: Philology, Old and New, in German Medieval Studies. *Digital Philology: A Journal of Medieval Cultures* 6(2): 177–195. https://doi.org/10.1353/dph.2017.0009

Björner, Erik Julius (1737) *Nordiska kämpa dater i en sagoflock samlade om forna kongar och hjältar. Volumen historicum, continens variorum in orbe hyperboreo antiquo regum, heroum* *et pugilum res praeclare et mirabiliter gestas. Accessit, praeter conspectum genealogicum Svethicorum regum et reginarum accuratissimum etiam praefatio.* Stockholm: J.L. Horrn.

Bloch, R. Howard (1990) New Philology and Old French. *Speculum* 65(1): 38–58. https://doi.org/10.2307/2864471

Bloom, Harold (1979) The Breaking of Form. In Harold Bloom (ed.) *De-construction and Criticism*: 1–38. New York: Seabury Press.

Busby, Keith (ed.) (1993) *Towards a Synthesis? Essays on the New Philology?* Amsterdam and Atlanta: Radopi. https://doi.org/10.1163/9789004650190

Cerquiglini, Bernard (1989) *Eloge de la variante, histoire critique de la philologie*. Paris: Seuil.

Driscoll, Matthew (2010) The Words on the Page: Thoughts on Philology, Old and New. In Judy Quinn and Emily Lethbridge (eds) *Creating the medieval saga: versions, variability, and editorial interpretations of Old Norse saga literature*: 87–104. Odense: University Press of Southern Denmark.

Eliot, Simon and Rose, Jonathan (eds) (2008) *A Companion to the History of the Book*. Oxford: Blackwell.

Eiríkur Jónsson (1892) *Hauksbók, udgiven efter de arnamagnæanske Håndskrifter No. 371, 544 og 675, 4° samt forskellige Papirshåndskrifter.* Copenhagen: Thieles Bogtrykkeri.

Felsenstein, Joseph (2013) *PHYLIP (Phylogeny Inference Package) 3.695*. Seattle: Department of Genome Sciences. <http://evolution.genetics.washington.edu/phylip/doc/main.html> (24 September 2023).

Felsenstein, Joseph (no date) *PHYLIP (Phylogeny Inference Package) version 3.6. Distributed by the author. Department of Genome Sciences, University of Washington, Seattle.* http://evolution.genetics.washington.edu/phylip/software.html (24 September 2023).

Finnur Jónsson (1900) *Landnámabók, Hauksbók, Sturlubók, Melabók m.m.* Copenhagen: Thieles Bogtrykkeri.

Frederiksen, Britta Olrik (1991) Det første stemma, dets videnskabshistoriske baggrund og skaber(e). In *The Eighth International Saga Conference. Preprints*: 110–20. Göteborg: Göteborgs Universitet.

Frederiksen, Britta Olrik (2003) Under stregen - lidt om det eksterne variantapparat i historisk perspektiv. In Pia Forssell *and* Rainer Knapas (eds) *Varianter och bibliografisk beskrivning*: 13–78. Ekenäs: Svenska litteratursällskapet i Finland.

Gabler, Hans Walter (2005) Textual Criticism. *The Johns Hopkins Guide to Literary Theory & Criticism*. <https://litguide.press.jhu.edu/cgi-bin/view.cgi?eid=250&query=textual%20criticism> (24 September 2023).

Gaskell, Philip (1972) *A New Introduction to Bibliography*. Oxford: Clarendon Press.

Glauser, Jürg and Tranter, Stephen N. (1990) Romances, Rímur, Chapbooks. Problems of Popular Literature in Late Medieval and Early Modern Scandinavia. *Parergon* 8(2): 37–52. https://doi.org/10.1353/pgn.1990.0055

Greetham, David C. (1992) Textual Scholarship. In Edward Finegan et al. (eds) *Introduction to Scholarship in Modern Languages and Literatures*: 103–137. New York: MLA Press.

Greetham, David C. (1994) *Textual Scholarship: An Introduction*. New York: Garland Publishing.

Greetham, David C. (2007) What is Textual Scholarship? In Simon Eliot and Jonathan Rose (eds) *A Companion to the History of the Book*: 21–32. Oxford: Blackwell. https://doi.org/10.1002/9780470690949.ch2

Grubmüller, Klaus *et al.* (1973) Spätmittelalterliche Prosaforschung. DGF-Forschergruppe-Programm am Seminar für deutsche Philologie der Universität Würzburg. *Jahrbuch für Internationale Germanistik* 5(1): 156–176.

Guðbrandur Vigfússon and Unger C.R. (eds) (1860–1868) *Flateyjarbók, en Samling af norske Konge-Sagaer med indskudte mindre Fortællinger om Begivenheder i og udenfor Norge, samt Annaler.* Oslo: Malling.

Guðni Jónsson (ed.) (1954–1959) *Fornaldarsögur Norðurlanda*. Reykjavík: Islendingasagnaútgáfan.

Guðni Jónsson and Bjarni Vilhjálmsson (eds) (1943–1944) *Fornaldarsögur Norðurlanda*. Reykjavík.

Guillaumin, Jean-Baptiste (2020) Criticisms of Digital Methods. In Philip Roelli (ed.) *Handbook of Stemmatology: History, Methodology, Digital Approaches*: 339–356. Berlin: De Gruyter.

Hall, Alaric and Parsons, Katelin (2013) Making Stemmas With Small Samples, and Digital Approaches to Publishing Them: Testing the Stemma of *Konráðs saga keisarasonar*. *Digital Medievalist* 9. <https://journal.digitalmedievalist.org/articles/10.16995/dm.51/> (24 September 2023). https://doi.org/10.16995/dm.51

Hansen, Anne Mette (2004a) *Den danske bønnebogstradition i materialfilologisk belysning*. PhD thesis. University of Copenhagen.

Hansen, Anne Mette (2004b) The Book as Artefact – ESTS colloquium, University of Copenhagen, 21st–23rd November 2003. *Editio* 18: 239–241.

Hansen, Anne Mette (2006) Senmiddelalderlige bønnebøger: kulturarv og andagtsbøger. *Studier i nordisk: foredrag og årsberetning 2004-2005*: 89-102.

Hansen, Anne Mette (2012) AM 421 12mo: An Artefactual Philological Study. *Care and Conservation of Manuscripts* 13: 1–16.

Hansen, Anne Mette (2017) The Body Language of Text: The Relationship Between the Textual Content and the Physical Appearance of Text in the Process of Transmission (Scholarly Editing). In Dorthe Duncker and Bettina Perregaard (eds) *Creativity and Continuity: Perspectives on the Dynamics of Language Conventionalisation*: 57–81. Copenhagen: U Press.

Haugen, Odd Einar (2010) Stitching the Text Together: Documentary and Eclectic Editions in Old Norse Philology. In Judy Quinn and Emily Lethbridge (eds) *Creating the Medieval Saga: Versions, Variability and Editorial Interpretations of Old Norse Saga Literature:* 39–65. Odense: University Press of Southern Denmark.

Haugen, Odd Einar (2013) Tekstkritikk og tekstfilologi. In Odd Einar Haugen (ed.) *Handbok i norrøn filologi*: 76–126. Bergen: Fagbokforlaget.

Hoenen, Armin (2020a) Software tools. In Philip Roelli (ed.) *Handbook of Stemmatology: History, Methodology, Digital Approaches*: 327–338. Berlin: De Gruyter.

Hoenen, Armin (2020b) The Stemma as a Computational Model. In Philip Roelli (ed.) *Handbook of Stemmatology: History, Methodology, Digital Approaches*: 226–241. Berlin: De Gruyter.

Hooper, A.G. (1930) *Hrómundar saga Greipssonar*. MA thesis. University of Leeds.

Hooper, A.G. (1932a) *Bragða Ǫlvis saga and rímur*. PhD thesis. University of Leeds.

Hooper, A.G. (1932b) *Bragða-Ǫlvis saga* Now First Edited. *Leeds Studies in English* 1: 42–54.

Hooper, A.G. (1934) Hrómundar saga Gripssonar and the Griplur. *Leeds Studies in English* 3: 51–56.

Kapitan, Katarzyna Anna (2017) A Choice of Relationship-revealing Variants for a Cladistic Analysis of Old Norse Texts: some methodological considerations. In Koraljka Golub and Marcelo Milrad (eds) *Extended Papers of the International Symposium on Digital Humanities*: 52–74. Växjö: CEUR.

Kapitan, Katarzyna Anna (2018) *Studies in the Transmission History of Hrómundar saga Greipssonar*. PhD thesis. University of Copenhagen.

Kapitan, Katarzyna Anna (2019) A Danish Collection of Old Norse Sagas: Material-Philological and Textual Studies of Acc. 61. In Katarzyna Anna Kapitan, Beeke Stegmann, and Seán Vrieland (eds) *From text to artefact. Studies in honour of Anne Mette Hansen*: 39–46. Leeds: Kismet Press.

Kapitan, Katarzyna Anna (2021a) Afterlife of a Lost Saga: A Hitherto Unknown Adaptation of the Lost Saga of Hrómundur Gripsson. *Saga-Book* 45: 59–90.

Kapitan, Katarzyna Anna (2021b) Hrómundur in Prose and Verse: On the Relationships Between Four Versions of the Story of Hrómundur Greipsson. *Gripla* 32: 257–288. https://doi.org/10.33112/gripla.32.10

Kapitan, Katarzyna Anna (2021c) Manuscripts Derived from Printed Editions in the Transmission History of Hrómundar saga Greipssonar. In Matthew Driscoll and Nioclás Mac Cathmhaoil (eds) *Hidden Harmonies: Manuscript and print on the North Atlantic fringe, 1500-1900*: 79–114. Copenhagen: Museum Tusculanum.

Kapitan, Katarzyna Anna (2022) When a King of Norway Became a King of Russia: Danish Historiography and Early Transmission and Reception of Hrómundar saga Greipssonar. *Scandinavian Studies* 94(3): 316–351. https://doi.org/10.5406/21638195.94.3.03

Kapitan, Katarzyna Anna (2024a) Lost but Not Forgotten: The Saga of Hrómundur and Its Manuscript Transmission. Oxford: Taylor Institution Library.

Kapitan, Katarzyna Anna (2024b) Lost but Not Forgotten: The Saga of Hrómundur and its Manuscript Transmission (Supplementary Materials). Oxford: Taylor Institution Library. http://dx.doi.org/10.5287/ora-449zgmvdx

Kjeldsen, Alex Speed (2013) *Filologiske studier i kongesagahåndskriftet Morkinskinna*. Copenhagen: Museum Tusculanum.

Kondrup, Johnny (2011) *Editionsfilologi*. Copenhagen: Museum Tusculanum.

Kölbing, Eugen (1876) *Beiträge zur Vergleichenden Geschichte der Romantischen Poesie und Prosa des Mittelalters*. Breslau: Verlag von Wilhelm Koebner.

Lavender, Philip (2014) *Whatever Happened to* Illuga saga Gríðfóstra*?: Origin, Transmission and Reception of a* Fornaldarsaga. PhD thesis. University of Copenhagen.

Lavender, Philip (2020) *Long Lives of Short Sagas: The Irrepressibility of Narrative and the Case of Illuga saga Gríðarfóstra*. Odense: University Press of Southern Denmark.

Leeuw van Weenen, Andrea de (1987) *Möðruvallabók, AM 132 fol*. Leiden: Brill

Leeuw van Weenen, Andrea de (2009) *Alexanders saga, AM 519a 4to in the Arnamagnæan Collection, Copenhagen.* Copenhagen: Museum Tusculanum.

Leslie, Helen (2012) Younger Icelandic Manuscripts and Old Norse Studies. *The Retrospective Methods Network Newsletter* 4: 148-161.

Löser, Freimut (2016) Überlieferungsgeschichte(n) schreiben. In *Überlieferungsgeschichte transdisziplinär*: 1–19. Wiesbaden: Dr. Ludwig Reichert Verlag.

Maas, Paul (1960) *Textkritik*. Leipzig: B.G. Teubner.

Macé, Caroline (2020) The Stemma as a Historical Tool. In Philip Roelli (ed.) *Handbook of Stemmatology: History, Methodology, Digital Approaches*: 272–291. Berlin: De Gruyter.

Maddox, Donald (1993) Philology: Philo-logos, Philo-logica or *Philologicon*? In Keith Busby (ed.) *Towards a synthesis? Essays on the New Philology?*: 59-70. Amsterdam: Radopi. https://doi.org/10.1163/9789004650190_006

Males, Mikael (2023) Textual Criticism and Old Norse Philology. *Studia Neophilologica*: 1–25. https://doi.org/10.1080/00393274.2023.2205888

McGann, Jerome John (1983) *A Critique of Modern Textual Criticism*. Charlottesville: University Press of Virginia.

McKenzie, Donald Francis (1986) *Bibliography and the Sociology of Texts*. London: The British Library.

Myrvoll, Klaus Johan (2023) The ideo-political background of 'new philology'. *Studia Neophilologica*: 1-7. https://doi.org/10.1080/00393274.2023.2228845

Nichols, Stephen (1988) Editor's Preface. *Romanic Review* 79(1): 1–3.

Nichols, Stephen (1990) Introduction: Philology in a Manuscript Culture. *Speculum* 65: 1–10. https://doi.org/10.2307/2864468

Nichols, Stephen (1994) Philology and Its Discontents. In *The Future of the Middle Ages*: 113–141. Gainesville: University Press of Florida.

Nichols, Stephen (1997) Why Material Philology? Some Thoughts. *Zeitschrift für deutsche Philologie* 116: 10–30.

Rafn, Carl Christian (ed.) (1829–1830) *Fornaldar sögur Nordrlanda eptir gömlum handritum*. Copenhagen: Popp.

Reynolds, L.D. and Wilson, N.G. (1974) Textual Criticism. In *Scribes and Scholars: A Guide to the Transmission of Greek and Latin Literature*: 186–213. Oxford: Clarendon Press.

Roelli, Philip (2020a) Definition of Stemma and Archetype. In Philip Roelli (ed.) *Handbook of Stemmatology: History, Methodology, Digital Approaches*: 209–225. Berlin: De Gruyter. https://doi.org/10.1515/9783110684384

Roelli, Philip (2020b) *Handbook of Stemmatology: History, Methodology, Digital Approaches*. Berlin: De Gruyter. https://doi.org/10.1515/9783110684384

Roos, Teemu (2020) Computational Construction of Trees. In Roelli, Philip (ed.) *Handbook of Stemmatology: History, Methodology, Digital Approaches*: 315-327. Berlin: De Gruyter.

Ruh, Kurt (1985a) Überlieferungsgeschichte mittelalterlicher Texte als methodischer Ansatz zu einer erweiterten Konzeption von Literaturgeschichte. In Kurt Ruh (ed.) *Überlieferungsgeschichtliche Prosaforschung: Beiträge der Würzburger Forschergruppe zur Methode und Auswertung*: 264–272. Tübingen: Max Niemeyer. https://doi.org/10.1515/9783110925883.262

Ruh, Kurt (ed.) (1985b) *Überlieferungsgeschichtliche Prosaforschung, Beiträge der Würzburger Forschergruppe zur Methode und Auswertung*. Tübingen: Max Niemeyer. https://doi.org/10.1515/9783110925883

Sandys, John Edwin (1903) *A History of Classical Scholarship. From the End of the Sixth Century B.C. to the End of the Middle Ages*. Cambridge: Cambridge University Press (reprint 2011). https://doi.org/10.1017/CBO9780511903717

Slay, Desmond (1997) *Mírmanns saga*. Copenhagen: C.A. Reitzel.

Stevenson, Angus and Waite, Maurice (eds) (2011) *Concise Oxford English Dictionary*. Oxford: Oxford University Press.

Sverrir Tómasson (2002) Er nýja textafræðin ný? *Gripla* 13: 199–216.

Tanselle, G. Thomas (1981) Textual Scholarship. In *Introduction to Scholarship in Modern Languages and Literatures*: 29–52. New York: The Modern Language Association of America.

Tanselle, G. Thomas (1992) *A Rationale of Textual Criticism*. Philadelphia: University of Pennsylvania Press. https://doi.org/10.9783/9780812200423

TEI Consortium (2020) TEI P5: Guidelines for Electronic Text Encoding and Interchange. 4.0.0. <http://www.tei-c.org/Guidelines/P5/> (18 March 2020).

Teresa Dröfn F. Njarðvík (2019) From the Margins into the Text: Material Influence on the Textuality of Bragða-Ölvis saga. A Conference presentation at the International Medieval Congress Leeds.

Thorpe, James (1972) *Principles of Textual Criticism*. San Marino: The Huntington Library.

Timpanaro, Sebastiano (2005) *The Genesis of Lachmann's Method*. Chicago: The University of Chicago Press.

Trovato, Paolo (2014) *Everything You Always Wanted to Know About Lachmann's Method: A Non-standard Handbook of Genealogical Textual Criticism in the Age of Post-structuralism, Cladistic, and Copy-text*. Padova: Libreriauniversitaria.it (Storie e linguaggi).

Unger, C.R. (1871) *Codex Frisianus, en Samling af norske Konge-Sagaer.* Oslo: Malling.

Valdimar Ásmundarson (ed.) (1886) *Fornaldarsögur Norðrlanda*. Reykjavík: Sigurður Kristjánsson.

Varvaro, Alberto (1999) The "New Philology" from an Italian Perspective. *Text* 12: 49–58.

Vitz, Evelyn Birge (1993) On the Role of a Renewed Philology in the Study of a Manuscript- and an Oral-culture. In Keith Busby (ed.) *Towards a synthesis? Essays on the New Philology:* 71–78. Amsterdam: Rodopi. https://doi.org/10.1163/9789004650190_007

Westra, Haijo J. (2014) What's in a Name: Old, New, and Material Philology, Textual Scholarship, and Ideology. In *Neo-Latin Philology: Old Tradition, New Approaches*: 13–24. Leuven: Leuven University Press. https://doi.org/10.2307/j.ctt14jxt24.4

Williams, Henrik (2009) Förnyad filologi. In *Omodernt: Människor och tankar i förmodern tid*: 272–292. Lund: Nordic Academic Press.

Williams-Krapp, Werner (2000) Die überlieferungsgeschichtliche Methode: Rückblick und Ausblick. *Internationales Archiv für Sozialgeschichte der deutschen Literatur* 25(2): 1–21. https://doi.org/10.1515/iasl.2000.25.2.1

Wolf, Kirsten (1993) Old Norse - New Philology. *Scandinavian Studies* 65: 338–348.

Zeevaert, Ludger *et al.* (2013) A New Stemma of *Njáls saga* - A Working Paper. <https://www.academia.edu/7317515/A_New_Stemma_of_Njáls_saga> (24 September 2023).

Zumthor, Paul (1972) Essai de poétique médiévale. Paris: Seuil.

Zundert, Joris van (2020) Computational Methods and Tools: Introductory Remarks. In Roelli, Philip (ed.) *Handbook of stemmatology: History, Methodology, Digital Approaches*: 292-294. Berlin: De Gruyter. https://doi.org/10.1515/9783110684384-006